\documentclass[twocolumn,amsmath,amssymb]{revtex4}

\usepackage{graphicx}
\usepackage{dcolumn}
\usepackage{bm}

\begin{document}

\title{The Harmonic Measure of Diffusion-Limited Aggregates including Rare Events}

\author{D. A. Adams$^{1}$, L. M. Sander$^{1,2}$, E. Somfai$^{4}$, R. M. Ziff $^{2,3,5}$}



\begin{abstract}
We obtain the harmonic measure of diffusion-limited aggregate (DLA) clusters  using a biased random-walk sampling technique which allows us to measure probabilities of random walkers hitting sections of clusters with unprecedented accuracy; our results include probabilities as small as $10^{-80}$. We find the multifractal $D(q)$ spectrum including regions of small and negative $q$. Our algorithm allows us to obtain the harmonic measure for clusters more than an order of magnitude larger than those achieved using the method of iterative conformal maps, which is the previous best method. We find a phase transition in the singularity spectrum $f(\alpha)$ at $\alpha \approx 14$ and also find a minimum $q$ of $D(q)$, $q_{min} = 0.9 \pm 0.05$.
\end{abstract}

\address{$^{1}$Department of Physics, University of Michigan, Ann Arbor MI 48109-1040 \\
$^{2}$Michigan Center for Theoretical Physics, University of Michigan, Ann Arbor MI 48109-1040 \\
$^{3}$Center for the Study of Complex Systems, University of Michigan, Ann Arbor Michigan, 48109-1040,  USA \\
$^{4}$Department of Physics and Centre for Complexity Science, University of Warwick, Coventry CV4 7AL, UK \\
$^{5}$Department of Chemical Engineering, University of Michigan, Ann Arbor MI 49109-2136}

\maketitle

\section{Introduction}

Diffusion-limited aggregation (DLA) is a stochastic model for irreversible growth which gives rise to fractal clusters \cite{Witten81,Sander00}, see Figures \ref{fig:Minkowski},\ref{fig:HarmonicMeasure}.   The growth process is defined by  releasing a random walker far from the cluster and allowing it to diffuse until it sticks to the surface and becomes part of the cluster. Then another particle is released, and so forth. The probability of sticking at various points on the cluster, i.e. the distribution of the growth probability,  is a function with very large variations. It is the subject of this paper. 

Since the Laplace equation is equivalent to the steady-state diffusion equation, this probability distribution is proportional to the perpendicular electric field on the surface of a charged electrode with the shape of the cluster; in this context the probability is called the  harmonic measure, and is defined for any surface.  For fractal surfaces, including that of DLA, the harmonic measure is usually multifractal \cite{Mandelbrot90}.  For DLA the harmonic measure is of particular interest because of the connection with the growth probability. For other fractal surfaces this connection is lost. However, the measure is still of substantial practical interest because its relationship with physical processes such as catalysis \cite{Grebenkov05}. 

For many interesting equilibrium fractals the harmonic measure can be calculated using conformal field theory \cite{Belikov08, Duplantier08, Bettelheim05} or Schramm-Loewner evolution (SLE) \cite{Gruzberg06}. There is no corresponding theory for DLA for which the measure must be found numerically. There are numerous studies in the literature of this quantity, for example \cite{Meakin86,Ball90,Jensen02,Hanan08}. This is a difficult problem because of the very large variation of the growth probability. As we will see the dynamic range of the function is of the order of $10^{80}$ even for rather small clusters. This is far out of the range accessible to straightforward random walker sampling.

In this paper, we use a biased random-walk sampling method. We can  obtain extremely small growth probabilities and  to accurately obtain the complete harmonic measure for DLA clusters of up to $10^6$ particles.  The method was previously used on percolation and Ising clusters \cite{Adams08}. For those (equilibrium) systems, we found good agreement with analytic predictions for the harmonic measure \cite{Duplantier99, Duplantier00}. 

The harmonic measure is usually characterized in terms of the generalized dimensions $D(q)$.  For integer $q$, $D(q)$ corresponds to the fractal dimension of the $q$ point correlation function.  We define  $D(q)$  by partitioning the external boundary of a DLA cluster into boxes of length $l$.  The probability that a diffusing particle will hit the section of the perimeter contained in box $i$ is denoted by $p_i$. These probabilities define a ``partition function"  $Z_l(q) = \sum_i (p_i)^q$ \cite{Halsey86}. If  $Z_l(q)$  can be written as a power law in the dimensionless ratio $R/l$, where $R$ is the overall size of the cluster, then the generalized dimension is given by:
\begin{equation}
\label{Zq}
Z_l(q) = (R/l)^{-\tau_q} = (R/l)^{-(q-1)D(q)}.
\end{equation}
There are special values of $D(q)$ including the box-counting dimension, $D(0) \approx 1.71$ \cite{TolmanMeakin89}. For two dimensional clusters we always have  $D(1) = 1$  \cite{Makarov}.  

Another quantity of interest is $f(\alpha)$, which is called the singularity spectrum. This function  is the Legendre transform of $\tau(q)$: 
\begin{equation}
\label{f_alpha}
f(\alpha) = q \frac{d \tau}{d q} - \tau, \quad \alpha = \frac{d \tau}{d q}.
\end{equation}
As is the case for $D(q)$, some special values of $f(\alpha)$ are known: $f(1) = 1$ and the largest value of $f(\alpha)$ is equal to $D(0)$.  
$f(\alpha)$ can have a phase transition, namely a maximum value of $\alpha$, $\alpha_{max}$, for which $f(\alpha)$  is defined. There has been significant disagreement as to whether $f(\alpha)$ for DLA has a phase transition.  This controversy is summarized in \cite{Jensen02}. 

The main difficulty in resolving this issue is that large $\alpha$, or small $q$, corresponds to the smallest probabilities on the cluster.  The straightforward method of obtaining the harmonic measure, sending large numbers of random walkers at the cluster, is only capable of measuring probabilities down to $\approx 10^{-10}$; even clusters with only $1,000$ particles have sections with growth probabilities significantly smaller than that. This issue was partially resolved by Jensen et al. \cite{Jensen02}, who used the method of iterated conformal maps (CM) \cite{Hastings98, Davidovitch00, Davidovitch01}, to obtain significantly smaller probabilities.  Their main result was the determination of the harmonic measure of a single cluster of size $3 \cdot 10^4$, where they found probabilities down to $10^{-35}$.  
This work \cite{Jensen02} was a significant advance, though the CM results are not conclusive in giving the asymptotic results for DLA because the CM method is limited to small clusters, and it is known that some features of DLA have slow crossover to asymptotic scaling \cite{Somfai99}.  There are other, technical, questions about the CM method that we discuss below.

Other groups have obtained the harmonic measure for on-lattice clusters using relaxation methods to solve the Laplace equation. Ball and Spivack \cite{Ball90} grew DLA clusters, corrected for lattice anisotropy,  up to $10^5$ particles. They then solved Laplace's equation numerically to obtain the measure. Hanan et al. \cite{Hanan08} measured the complete harmonic measure of DLA clusters using a related relaxation technique. In contrast to  \cite{Ball90}, these authors first grew the cluster off-lattice, then forced it on lattice to solve for the measure. The simulations in \cite{Hanan08} were also limited to small clusters of  $6 \cdot 10^4$ particles.

\section{Simulation Methods}

We grow our DLA clusters by the method that is now standard \cite{Sander00}, which includes speeding up the process by allowing the random walker particles to take large jumps. We store the  cluster in a data structure which allows the calculation of the size of the jump to be performed in $O(\log(n))$ time, where $n$ is the cluster size. These methods allow us to grow clusters in $O(n \log(n))$ time, a big improvement over the CM method, which is $O(n^2)$. In the CM method the harmonic measure is available at each step. In our case, we use a biased random walker method (the signpost method \cite{Adams08}) to obtain the harmonic measure once the cluster has finished growing.  

\begin{figure}
\includegraphics[width=0.45\textwidth]{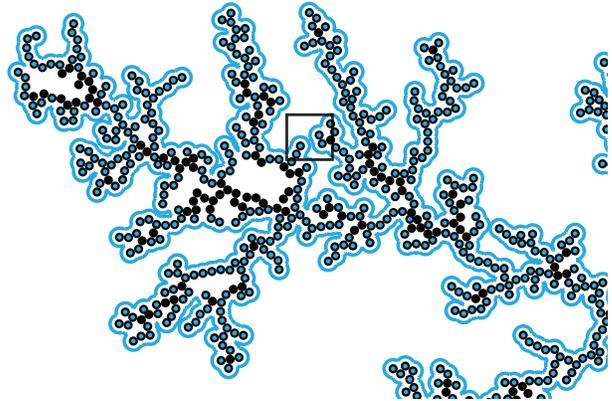}
\caption{\label{fig:Minkowski} A branch of a DLA cluster with an external border representing the Minkowski cover of the cluster. The particles filled grey (blue online) in the center are accessible to random walkers. The solid black particles can never be reached. Inside the gray box is a narrow neck which marks a low probability pathway for random walkers.}
\end{figure}

The signpost method consists of  two iterated steps: a sampling step and a measurement step.  On the first step, a large number of random walker probe particles ($N$), each with weight $1/N$, are released far from the cluster and diffuse until they hit the cluster.  This allows us to determine which areas of the cluster are poorly sampled.  Next, we place line segments (signposts) blocking off all regions the cluster that have sites that are hit by fewer than some percentage of the probe particles, say $10\%$. In the measurement step, we release $N$ more probe particles far from the cluster and allow them to hit the cluster and signposts. The probe particles that hit the cluster have their weight permanently added to the perimeter site probability distribution.  The locations on the signposts where the probe particles hit in the first step are used as the initial location of the $N$ probe particles released in the probe step of the second iteration. To conserve probability, each probe particle released in the second iteration has weight $p / N$, where $p$ is the fraction of probe particles that absorbed on the signpost in the first iteration.  The probe particles released in the probe step of the second iteration help determine which sections of the cluster are still poorly sampled.  More signposts are added to block off the still poorly sampled regions and then the probe particles for the measurement step are released. This process is repeated until the growth probability of all sites has been measured.  For a more detailed description of the algorithm, see \cite{Adams08}. 

This method is similar to a rare event method in chemical physics that uses `milestones' \cite{Chem_1}.  The main difference between the our method and that of \cite{Chem_1} is we choose the locations of the signposts/milestones \emph{ dynamically} and that we do not need to know an \emph{a priori} distribution for the random walkers along the milestones. See also \cite{Allen06}.

\begin{figure}
\includegraphics[width=0.45\textwidth]{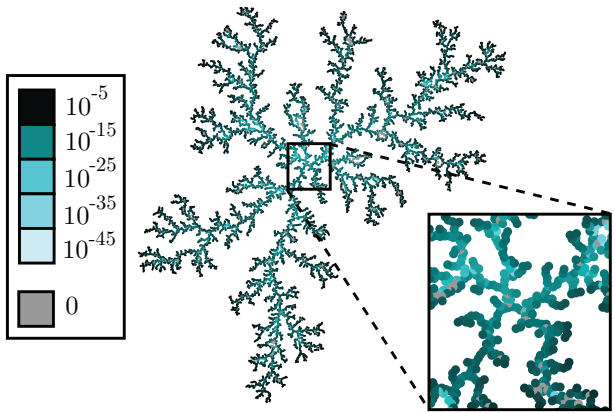}
\caption{\label{fig:HarmonicMeasure} The Harmonic Measure for a cluster with $10^4$ particles, the lighter the color the smaller the measure. The size of the particles is doubled to represent the cover of the DLA cluster. The smallest measure in the cluster is $\approx 10^{-49}$. Sites that cannot be reached are marked grey. }
\end{figure}

This signpost method  allowed us to measure probabilities down to $10^{-300}$ for percolation and Ising clusters.  For DLA we have measured probabilities down to $10^{-80}$.  Performing the signpost algorithm on DLA clusters is more complex than the percolation and Ising cases. DLA clusters are grown off-lattice, which means that some sections of the exterior of the cluster are almost completely blocked by two branches of the cluster nearly meeting, making a narrow passage slightly larger than the diameter of a probe particle; see Figure \ref{fig:Minkowski}.  The probability of a probe particle diffusing through some of these passages without touching the cluster is smaller than $10^{-8}$. These passages are treated differently, but in a way consistent with the signpost algorithm. Specifically, we slowly move signposts perpendicular to the passage inward over several iterations until the probe particles can reach other sections of perimeter. In other words, we allow narrow passages to have closely spaced signpost lines so that proper sampling can be achieved.

One minor difference between the signposting we use in this paper and the one used previously is that we now reduce the threshold for blocking off sections of the cluster as a function of the total number of probe particles that hit signposts in the previous iteration. When more probe particles hit the signposts, we move the signposts much deeper the next time. Previously, we reduced the threshold a fixed amount each iteration.  We found that this dynamic threshold adjustment gave us more consistent particle saturation on the signposts, which in turn decreased the rate at which the error grew from step to step.

Before we can apply the signpost method to a DLA cluster, we first must find the perimeter of the cluster. More precisely, we must find all sites that are accessible to the probe particles.  First, we take a ball the size of a probe particle and roll it clockwise around the cluster particle furthest from the center of the cluster until the ball touches a second cluster particle.  After that, the ball is rolled clockwise about the second cluster particle until it touches a third particle.  This process is repeated until the ball returns to its initial location.  Note that a single cluster particle can be visited more than once by the ball.   This process finds something akin to the Minkowski cover of the cluster, Figure \ref{fig:Minkowski}. We found that on average only $80$\% of the particles in cluster are accessible to random walkers.  This means that $20$\% of the particles have a measure of exactly zero and these regions of the cluster will never grow.  We found the $80$\% accessibility to be constant over a range of large cluster sizes, which shows that the accessible perimeter has the same fractal dimension of the complete perimeter and the cluster itself, in contrast to percolation where the corresponding accessible perimeter has a smaller fractal dimension than the complete perimeter \cite{Saleur87, Duplantier99}. 

Our perimeter accessibility results agree qualitatively with other work \cite{Menshutin08}, which looked at cluster particle accessibility as a function of probe particle size.  However, the agreement is not precise because the authors of \cite{Menshutin08} measured the accessible cluster using $10^5$ random walkers, which are extremely unlikely to hit the low measure sections of the perimeter.  

\begin{figure}
\includegraphics[width=0.45\textwidth]{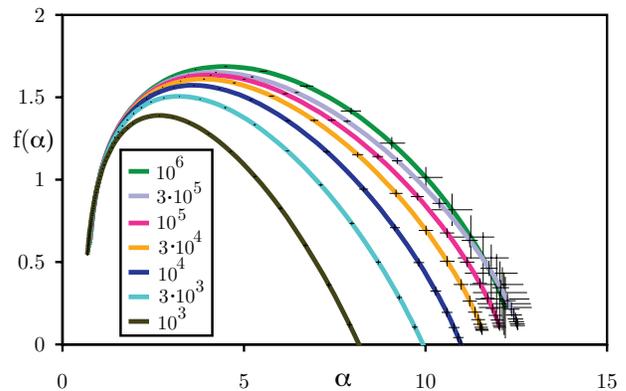}
\caption{\label{fig:Falpha} $f(\alpha)$ vs. $\alpha$ for seven different system sizes with error bars. Note that the spectra appear to be converging to some asymptotic spectrum. }
\end{figure}

We grew DLA clusters of various sizes: $10^3$, $3 \cdot 10^3$, $10^4$, $3 \cdot 10^4$, $10^5$, $3 \cdot 10^5$, and $10^6$ particles.  For each DLA cluster grown, we obtained the harmonic measure using the signpost algorithm. Figure \ref{fig:HarmonicMeasure} shows the harmonic measure for a cluster with $10^4$ particles.  The different cluster sizes required a different number of random walkers per iteration, $10^6$, $10^6$, $5 \cdot 10^6$, $10^7$, $2.5 \cdot 10^7$, $10^8$, and $2.5 \cdot 10^8$ for $10^3$, $3 \cdot 10^3$, $10^4$, $3 \cdot 10^4$, $10^5$, $3 \cdot 10^5$, and $10^6$ sized clusters, respectively.  The number of random walkers needed was estimated by determining the number of walkers required to get at least $10^4$ random walkers absorbed on each signpost for every iteration. We believe this is a conservative criterion. 

\begin{figure}
\includegraphics[width=0.45\textwidth]{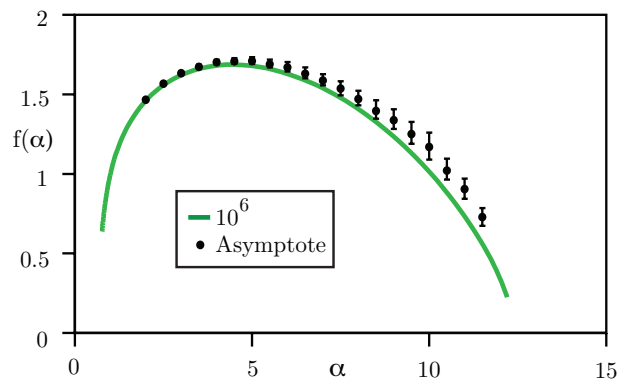}
\caption{\label{fig:FalphaAsym} $f(\alpha)$ vs. $\alpha$ for $n=10^6$ and the asymptotic estimate, dotted, with error bars estimated from the data in Figure \ref{fig:Falpha}.  Note that the asymptotic estimate terminates at $\alpha = 11$ only because there were too few system sizes to extrapolate for larger $\alpha$. We believe the phase-transition in $f(\alpha)$ occurs at $\alpha \approx 14$.}
\end{figure}

\section{Results}

We use the  method described above to obtain $D(q)$. First, we take the space that contains a cluster and section it into boxes of size $l$ and then measure $Z_l(q)$. We do this measurement of $Z_l(q)$ for various values of $l$ for a given $q$.  Next, we calculate the slope of the function $\ln Z_l(q)$ versus $\ln l$; this is $\tau(q)$, which when divided by $(q-1)$, gives $D(q)$.  The fit is performed over the range of $l$ for which the $\log$-$\log$ plot is linear.  This range is about one order of magnitude for the smallest system size and larger than one order of magnitude for larger systems.  With this set of $D(q)$'s for individual clusters of various sizes, we can calculate the average values of $D(q)$ for various sizes. We found that our results for large systems are close to the known values for $D(0)$ and $D(1)$, 1.66 and 0.99 respectively.

\begin{figure}
\includegraphics[width=0.45\textwidth]{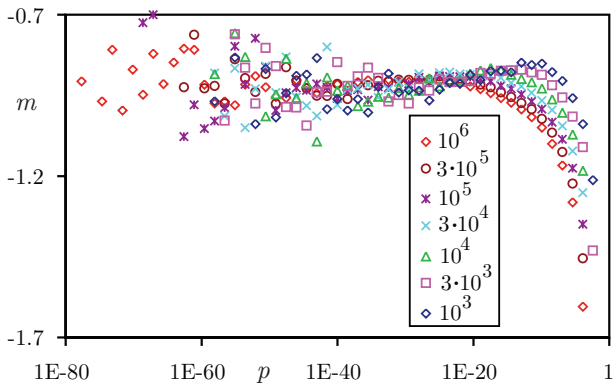}
\caption{\label{fig:hist} The slope of the power-law fit to the probability distribution at various points.  The slope at each probability is averaged over about an order of magnitude in probability.  }
\end{figure}

Using the results of $D(q)$ for individual clusters, we can Legendre transform the results to obtain $f(\alpha)$ for each cluster; see figure ~\ref{fig:Falpha}.  For a range of $\alpha$'s, we estimated the asymptotic  value of $f(\alpha)$ using finite-size scaling techniques in $n$, so that the correction to $f(\alpha)$ is of the form $n^{-\beta}$ where $\beta$ is a crossover exponent.  The asymptotic values were determined by minimizing the residual of the power-law fit, see  Fig.~\ref{fig:FalphaAsym}. We found best fit for $\beta$ was $0.4$ for the entire range of $\alpha$. Note that exponents of $1/3$ and $1/2$ are also consistent with the data.  This means we are consistent with \cite{Somfai99} where a crossover exponent of $1/3$ was found (for different quantities). The asymptotic $f(\alpha)$ values are consistent with special known values $f(1)$ and the maximum $f$, measured to be $1.00$ and $1.71$ respectively.  We believe the asymptotic $f(\alpha)$ calculated is the true $f(\alpha)$ for DLA. The last $\alpha$ for which $f(\alpha)$ is defined is more difficult to estimate. From visual inspection the asymptotic point of the phase transition appears to be about $\alpha \approx 14$. This is significantly smaller then the value found by the authors of \cite{Jensen02}, $\alpha \approx 18$. 

We were also able to obtain a histogram of the growth probabilities for every system size. The bins were sized logarithmically, to allow for a power-law fit to the results. Figure \ref{fig:hist} shows the slope of the power-law fit to the probability distribution.  The slope is fairly consistently $-0.9$ which corresponds to a smallest $q$ for which $D(q)$ is defined being $-0.1$.  These values agree moderately well with \cite{Jensen02}. 

\section{Conclusions}

In this paper we applied a rare-event technique to obtain the complete harmonic measure of DLA clusters. We found that probability distribution is consistent with a power-law exponent of -$0.9$. We also found a slow crossover to infinite-size cluster behavior in $f(\alpha)$, in agreement with previous work \cite{Somfai99, Hanan08}. We believe that our extrapolated  $f(\alpha)$ is a very good approximation to $f(\alpha)$ for infinite-sized DLA. We found a phase transition in $f(\alpha)$ at $\alpha \approx 14$. This maximum $\alpha$ is related to the opening angle of the branches near the seed point of the cluster. The area around the seed point should have the lowest measure, so the angles in that region, $\phi_{max}$ are related to the largest alpha by $\alpha_{max} \sim 1/ \phi_{max}$, \cite{Jensen02}. By determining the exact relationship between $\alpha_{max}$ and $\phi_{max}$, Hanan and Heffernan \cite{Hanan08} determined the asymptotic $\alpha_{max}$ as $\alpha_{max} \approx 15$ using results from Mandelbrot et al. \cite{Mandelbrot02} for the asymptotic estimate of $\phi_{max}$. This is in satisfactory agreement with our results for $\alpha_{max}$. 

Our results differ significantly from those obtained by the CM method \cite{Jensen02} in several ways. First, we find a significantly smaller value for $\alpha_{max}$. Second, we find that finite-size effects are still noticeable on clusters with $10^6$ particles. The authors of \cite{Jensen02} found no finite-size effects at their largest system size, $3 \cdot 10^4$ particles. This is inconsistent with our findings. Lastly, we find that the smallest probabilities found on clusters of size $3 \cdot 10^4$ are significantly smaller, about $5$ to $10$ orders of magnitude, than reported for CM clusters. We do find good agreement for small and moderate values of $\alpha$, which corresponds to region of the spectrum which is easily measured by random walker sampling. This explains why the difference between CM and standard DLA clusters was not seen earlier.

Assuming that the signpost method and the CM method are both successful at obtaining the measure for their respective clusters, then the only explanation for the discrepancy is that CM clusters are not the same as DLA clusters grown using particles.  Superficially, CM clusters appear to be the cover of DLA clusters.   If this were the case, then both methods would obtain the same measure because the measure for a probe particle the same size as a cluster particle hitting a standard cluster is exactly the same as the measure for a point-sized particle hitting the cover of the same cluster.   The heart of the issue may be the shape and size of the `bumps' added to the CM clusters during each step.  These bumps are designed to have a semicircular shape, and to be of fixed size, but they can distort as noted in \cite{Stepanov01, Somfai03}. It is important to check that the bumps are, in fact of fixed size, and resize them if necessary. It is  not clear that this was done in  \cite{Jensen02}. Even if this correction were made, the shape of the bumps can be very distorted deep inside the cluster.



\acknowledgments
We would like to thank Robin Ball for useful discussions. This work was supported in part by the National Science Foundation through DMS-0553487.

\bibliography{DLA_Harmonic_Draft_arXiv}

\begin{thebibliography}{29}
\expandafter\ifx\csname natexlab\endcsname\relax\def\natexlab#1{#1}\fi
\expandafter\ifx\csname bibnamefont\endcsname\relax
  \def\bibnamefont#1{#1}\fi
\expandafter\ifx\csname bibfnamefont\endcsname\relax
  \def\bibfnamefont#1{#1}\fi
\expandafter\ifx\csname citenamefont\endcsname\relax
  \def\citenamefont#1{#1}\fi
\expandafter\ifx\csname url\endcsname\relax
  \def\url#1{\texttt{#1}}\fi
\expandafter\ifx\csname urlprefix\endcsname\relax\def\urlprefix{URL }\fi
\providecommand{\bibinfo}[2]{#2}
\providecommand{\eprint}[2][]{\url{#2}}

\bibitem[{\citenamefont{Witten and Sander}(1981)}]{Witten81}
\bibinfo{author}{\bibfnamefont{T.~A.} \bibnamefont{Witten}} \bibnamefont{and}
  \bibinfo{author}{\bibfnamefont{L.~M.} \bibnamefont{Sander}},
  \bibinfo{journal}{Phys. Rev. Lett.} \textbf{\bibinfo{volume}{47}},
  \bibinfo{pages}{1400} (\bibinfo{year}{1981}).

\bibitem[{\citenamefont{Sander}(2000)}]{Sander00}
\bibinfo{author}{\bibfnamefont{L.~M.} \bibnamefont{Sander}},
  \bibinfo{journal}{Contemporary Physics} \textbf{\bibinfo{volume}{41}},
  \bibinfo{pages}{203} (\bibinfo{year}{2000}).

\bibitem[{\citenamefont{Mandelbrot and Evertsz}(1990)}]{Mandelbrot90}
\bibinfo{author}{\bibfnamefont{B.~B.} \bibnamefont{Mandelbrot}}
  \bibnamefont{and} \bibinfo{author}{\bibfnamefont{C.~J.~G.}
  \bibnamefont{Evertsz}}, \bibinfo{journal}{Nature}
  \textbf{\bibinfo{volume}{348}}, \bibinfo{pages}{143} (\bibinfo{year}{1990}).

\bibitem[{\citenamefont{Grebenkov}(2005)}]{Grebenkov05}
\bibinfo{author}{\bibfnamefont{D.~S.} \bibnamefont{Grebenkov}},
  \bibinfo{journal}{Phys. Rev. Lett.} \textbf{\bibinfo{volume}{95}},
  \bibinfo{pages}{200602} (\bibinfo{year}{2005}).

\bibitem[{\citenamefont{Belikov et~al.}(2008)\citenamefont{Belikov, Gruzberg,
  and Rushkin}}]{Belikov08}
\bibinfo{author}{\bibfnamefont{A.}~\bibnamefont{Belikov}},
  \bibinfo{author}{\bibfnamefont{I.~A.} \bibnamefont{Gruzberg}},
  \bibnamefont{and} \bibinfo{author}{\bibfnamefont{I.}~\bibnamefont{Rushkin}},
  \bibinfo{journal}{J. Phys. A} \textbf{\bibinfo{volume}{41}},
  \bibinfo{pages}{285006} (\bibinfo{year}{2008}).

\bibitem[{\citenamefont{Duplantier and Binder}(2008)}]{Duplantier08}
\bibinfo{author}{\bibfnamefont{B.}~\bibnamefont{Duplantier}} \bibnamefont{and}
  \bibinfo{author}{\bibfnamefont{I.~A.} \bibnamefont{Binder}},
  \bibinfo{journal}{Nucl. Phys. B} \textbf{\bibinfo{volume}{802}},
  \bibinfo{pages}{494} (\bibinfo{year}{2008}).

\bibitem[{\citenamefont{Bettelhiem et~al.}(2005)\citenamefont{Bettelhiem,
  Rushkin, Gruzberg, and Wiegmann}}]{Bettelheim05}
\bibinfo{author}{\bibfnamefont{E.}~\bibnamefont{Bettelhiem}},
  \bibinfo{author}{\bibfnamefont{I.}~\bibnamefont{Rushkin}},
  \bibinfo{author}{\bibfnamefont{I.~A.} \bibnamefont{Gruzberg}},
  \bibnamefont{and} \bibinfo{author}{\bibfnamefont{P.}~\bibnamefont{Wiegmann}},
  \bibinfo{journal}{Phys. Rev. Lett.} \textbf{\bibinfo{volume}{95}},
  \bibinfo{pages}{170602} (\bibinfo{year}{2005}).

\bibitem[{\citenamefont{Gruzberg}(2006)}]{Gruzberg06}
\bibinfo{author}{\bibfnamefont{I.~A.} \bibnamefont{Gruzberg}},
  \bibinfo{journal}{J. Phys. A} \textbf{\bibinfo{volume}{39}},
  \bibinfo{pages}{12601} (\bibinfo{year}{2006}).

\bibitem[{\citenamefont{Meakin et~al.}(1986)\citenamefont{Meakin, Coniglio,
  Stanley, and Witten}}]{Meakin86}
\bibinfo{author}{\bibfnamefont{P.}~\bibnamefont{Meakin}},
  \bibinfo{author}{\bibfnamefont{A.}~\bibnamefont{Coniglio}},
  \bibinfo{author}{\bibfnamefont{H.~E.} \bibnamefont{Stanley}},
  \bibnamefont{and} \bibinfo{author}{\bibfnamefont{T.~A.}
  \bibnamefont{Witten}}, \bibinfo{journal}{Phys. Rev. A}
  \textbf{\bibinfo{volume}{34}}, \bibinfo{pages}{3325} (\bibinfo{year}{1986}).

\bibitem[{\citenamefont{Ball and Spivack}(1990)}]{Ball90}
\bibinfo{author}{\bibfnamefont{R.~C.} \bibnamefont{Ball}} \bibnamefont{and}
  \bibinfo{author}{\bibfnamefont{O.~R.} \bibnamefont{Spivack}},
  \bibinfo{journal}{J. Phys. A} \textbf{\bibinfo{volume}{23}},
  \bibinfo{pages}{5295} (\bibinfo{year}{1990}).

\bibitem[{\citenamefont{Jensen et~al.}(2002)\citenamefont{Jensen, Levermann,
  Mathiesen, and Procaccia}}]{Jensen02}
\bibinfo{author}{\bibfnamefont{M.~H.} \bibnamefont{Jensen}},
  \bibinfo{author}{\bibfnamefont{A.}~\bibnamefont{Levermann}},
  \bibinfo{author}{\bibfnamefont{J.}~\bibnamefont{Mathiesen}},
  \bibnamefont{and}
  \bibinfo{author}{\bibfnamefont{I.}~\bibnamefont{Procaccia}},
  \bibinfo{journal}{Phys. Rev. E} \textbf{\bibinfo{volume}{65}},
  \bibinfo{pages}{046109} (\bibinfo{year}{2002}).

\bibitem[{\citenamefont{Hanan and Heffernan}(2008)}]{Hanan08}
\bibinfo{author}{\bibfnamefont{W.~G.} \bibnamefont{Hanan}} \bibnamefont{and}
  \bibinfo{author}{\bibfnamefont{D.~M.} \bibnamefont{Heffernan}},
  \bibinfo{journal}{Phys. Rev. E} \textbf{\bibinfo{volume}{77}},
  \bibinfo{pages}{011405} (\bibinfo{year}{2008}).

\bibitem[{\citenamefont{Adams et~al.}(2008)\citenamefont{Adams, Sander, and
  Ziff}}]{Adams08}
\bibinfo{author}{\bibfnamefont{D.~A.} \bibnamefont{Adams}},
  \bibinfo{author}{\bibfnamefont{L.~M.} \bibnamefont{Sander}},
  \bibnamefont{and} \bibinfo{author}{\bibfnamefont{R.~M.} \bibnamefont{Ziff}},
  \bibinfo{journal}{Phys. Rev. Lett.} \textbf{\bibinfo{volume}{101}},
  \bibinfo{pages}{144102} (\bibinfo{year}{2008}).

\bibitem[{\citenamefont{Duplantier}(1999)}]{Duplantier99}
\bibinfo{author}{\bibfnamefont{B.}~\bibnamefont{Duplantier}},
  \bibinfo{journal}{Phys. Rev. Lett.} \textbf{\bibinfo{volume}{82}},
  \bibinfo{pages}{3940} (\bibinfo{year}{1999}).

\bibitem[{\citenamefont{Duplantier}(2000)}]{Duplantier00}
\bibinfo{author}{\bibfnamefont{B.}~\bibnamefont{Duplantier}},
  \bibinfo{journal}{Phys. Rev. Lett.} \textbf{\bibinfo{volume}{84}},
  \bibinfo{pages}{1363} (\bibinfo{year}{2000}).

\bibitem[{\citenamefont{Halsey et~al.}(1986)\citenamefont{Halsey, Jensen,
  Kadanoff, Procaccia, and Shraiman}}]{Halsey86}
\bibinfo{author}{\bibfnamefont{T.~C.} \bibnamefont{Halsey}},
  \bibinfo{author}{\bibfnamefont{M.~H.} \bibnamefont{Jensen}},
  \bibinfo{author}{\bibfnamefont{L.~P.} \bibnamefont{Kadanoff}},
  \bibinfo{author}{\bibfnamefont{I.}~\bibnamefont{Procaccia}},
  \bibnamefont{and} \bibinfo{author}{\bibfnamefont{B.~I.}
  \bibnamefont{Shraiman}}, \bibinfo{journal}{Phys. Rev. A}
  \textbf{\bibinfo{volume}{33}}, \bibinfo{pages}{1141 } (\bibinfo{year}{1986}).

\bibitem[{\citenamefont{Meakin and Tolman}(1989)}]{TolmanMeakin89}
\bibinfo{author}{\bibfnamefont{P.}~\bibnamefont{Meakin}} \bibnamefont{and}
  \bibinfo{author}{\bibfnamefont{S.}~\bibnamefont{Tolman}},
  \bibinfo{journal}{Phys. Rev. A} \textbf{\bibinfo{volume}{40}},
  \bibinfo{pages}{428} (\bibinfo{year}{1989}).

\bibitem[{\citenamefont{Makarov}(1985)}]{Makarov}
\bibinfo{author}{\bibfnamefont{N.}~\bibnamefont{Makarov}},
  \bibinfo{journal}{Proc. London Math. Soc.} \textbf{\bibinfo{volume}{51}},
  \bibinfo{pages}{369} (\bibinfo{year}{1985}).

\bibitem[{\citenamefont{Hastings and Levitov}(1998)}]{Hastings98}
\bibinfo{author}{\bibfnamefont{M.~B.} \bibnamefont{Hastings}} \bibnamefont{and}
  \bibinfo{author}{\bibfnamefont{L.~S.} \bibnamefont{Levitov}},
  \bibinfo{journal}{Physica D} \textbf{\bibinfo{volume}{116}},
  \bibinfo{pages}{244} (\bibinfo{year}{1998}).

\bibitem[{\citenamefont{Davidovitch et~al.}(2000)\citenamefont{Davidovitch,
  Levermann, and Procaccia}}]{Davidovitch00}
\bibinfo{author}{\bibfnamefont{B.}~\bibnamefont{Davidovitch}},
  \bibinfo{author}{\bibfnamefont{A.}~\bibnamefont{Levermann}},
  \bibnamefont{and}
  \bibinfo{author}{\bibfnamefont{I.}~\bibnamefont{Procaccia}},
  \bibinfo{journal}{Phys. Rev. E} \textbf{\bibinfo{volume}{62}},
  \bibinfo{pages}{R5919} (\bibinfo{year}{2000}).

\bibitem[{\citenamefont{Davidovitch et~al.}(2001)\citenamefont{Davidovitch,
  Jensen, Levermann, Mathiesen, and Procaccia}}]{Davidovitch01}
\bibinfo{author}{\bibfnamefont{B.}~\bibnamefont{Davidovitch}},
  \bibinfo{author}{\bibfnamefont{M.~H.} \bibnamefont{Jensen}},
  \bibinfo{author}{\bibfnamefont{A.}~\bibnamefont{Levermann}},
  \bibinfo{author}{\bibfnamefont{J.}~\bibnamefont{Mathiesen}},
  \bibnamefont{and}
  \bibinfo{author}{\bibfnamefont{I.}~\bibnamefont{Procaccia}},
  \bibinfo{journal}{Phys. Rev. Lett.} \textbf{\bibinfo{volume}{87}},
  \bibinfo{pages}{164101} (\bibinfo{year}{2001}).

\bibitem[{\citenamefont{Somfai et~al.}(1999)\citenamefont{Somfai, Sander, and
  Ball}}]{Somfai99}
\bibinfo{author}{\bibfnamefont{E.}~\bibnamefont{Somfai}},
  \bibinfo{author}{\bibfnamefont{L.~M.} \bibnamefont{Sander}},
  \bibnamefont{and} \bibinfo{author}{\bibfnamefont{R.~C.} \bibnamefont{Ball}},
  \bibinfo{journal}{Phys. Rev. Lett.} \textbf{\bibinfo{volume}{83}},
  \bibinfo{pages}{26} (\bibinfo{year}{1999}).

\bibitem[{\citenamefont{Faradjian and Elber}(2004)}]{Chem_1}
\bibinfo{author}{\bibfnamefont{A.~K.} \bibnamefont{Faradjian}}
  \bibnamefont{and} \bibinfo{author}{\bibfnamefont{R.}~\bibnamefont{Elber}},
  \bibinfo{journal}{J. Chem. Phys.} \textbf{\bibinfo{volume}{120}},
  \bibinfo{pages}{10880} (\bibinfo{year}{2004}).

\bibitem[{\citenamefont{Allen et~al.}(2006)\citenamefont{Allen, Frenkel, and
  ten Wolde}}]{Allen06}
\bibinfo{author}{\bibfnamefont{R.~J.} \bibnamefont{Allen}},
  \bibinfo{author}{\bibfnamefont{D.}~\bibnamefont{Frenkel}}, \bibnamefont{and}
  \bibinfo{author}{\bibfnamefont{P.~R.} \bibnamefont{ten Wolde}},
  \bibinfo{journal}{J. Chem. Phys.} \textbf{\bibinfo{volume}{124}},
  \bibinfo{eid}{024102} (pages~\bibinfo{numpages}{16}) (\bibinfo{year}{2006}).

\bibitem[{\citenamefont{Saleur and Duplantier}(1987)}]{Saleur87}
\bibinfo{author}{\bibfnamefont{H.}~\bibnamefont{Saleur}} \bibnamefont{and}
  \bibinfo{author}{\bibfnamefont{B.}~\bibnamefont{Duplantier}},
  \bibinfo{journal}{Phys. Rev. Lett.} \textbf{\bibinfo{volume}{58}},
  \bibinfo{pages}{2325} (\bibinfo{year}{1987}).

\bibitem[{\citenamefont{Menshutin et~al.}(2008)\citenamefont{Menshutin, Shchur,
  and Vinokour}}]{Menshutin08}
\bibinfo{author}{\bibfnamefont{A.~Y.} \bibnamefont{Menshutin}},
  \bibinfo{author}{\bibfnamefont{L.~N.} \bibnamefont{Shchur}},
  \bibnamefont{and} \bibinfo{author}{\bibfnamefont{V.~M.}
  \bibnamefont{Vinokour}}, \bibinfo{journal}{Physica A}
  \textbf{\bibinfo{volume}{387}}, \bibinfo{pages}{6299} (\bibinfo{year}{2008}).

\bibitem[{\citenamefont{Mandelbrot et~al.}(2002)\citenamefont{Mandelbrot, Kol,
  and Aharony}}]{Mandelbrot02}
\bibinfo{author}{\bibfnamefont{B.~B.} \bibnamefont{Mandelbrot}},
  \bibinfo{author}{\bibfnamefont{B.}~\bibnamefont{Kol}}, \bibnamefont{and}
  \bibinfo{author}{\bibfnamefont{A.}~\bibnamefont{Aharony}},
  \bibinfo{journal}{Phys. Rev. Lett.} \textbf{\bibinfo{volume}{88}},
  \bibinfo{pages}{055501} (\bibinfo{year}{2002}).

\bibitem[{\citenamefont{Stepanov and Levitov}(2001)}]{Stepanov01}
\bibinfo{author}{\bibfnamefont{M.~G.} \bibnamefont{Stepanov}} \bibnamefont{and}
  \bibinfo{author}{\bibfnamefont{L.~S.} \bibnamefont{Levitov}},
  \bibinfo{journal}{Phys. Rev. E} \textbf{\bibinfo{volume}{63}},
  \bibinfo{pages}{061102} (\bibinfo{year}{2001}).

\bibitem[{\citenamefont{Somfai et~al.}(2003)\citenamefont{Somfai, Ball, DeVita,
  and Sander}}]{Somfai03}
\bibinfo{author}{\bibfnamefont{E.}~\bibnamefont{Somfai}},
  \bibinfo{author}{\bibfnamefont{R.~C.} \bibnamefont{Ball}},
  \bibinfo{author}{\bibfnamefont{J.~P.} \bibnamefont{DeVita}},
  \bibnamefont{and} \bibinfo{author}{\bibfnamefont{L.~M.}
  \bibnamefont{Sander}}, \bibinfo{journal}{Phys. Rev. E}
  \textbf{\bibinfo{volume}{68}}, \bibinfo{pages}{020401}
  (\bibinfo{year}{2003}).

\end{thebibliography}

\end{document}